\documentclass[reqno]{amsart}
\usepackage{amsmath}
\usepackage{amssymb}
  \usepackage{paralist} 
  \usepackage{graphics} 
  \usepackage{epsfig} 
\usepackage{graphicx}  \usepackage{epstopdf}
 \usepackage[colorlinks=true]{hyperref}

\hypersetup{urlcolor=blue, citecolor=red}

\theoremstyle{definition}

\newtheorem{remark}{Remark}

\usepackage{epstopdf}
\graphicspath{{./figures/}}

\usepackage[norelsize]{algorithm2e}

\newcommand{\putfig}[4]{ 
\begin{figure}[t!]  
\centering 
\includegraphics[width=#1\textwidth]{#2} 
\caption{#3}  
\label{#4} 
\end{figure} 
}

\title[Estimation of the social force parameters in crowd dynamics] 
      { Parameter estimation of social forces in crowd dynamics models via a probabilistic method}

\author[A. Corbetta, A. Muntean, F. Toschi, and K. Vafayi]{}



\begin{document}
\maketitle

\centerline{\scshape Alessandro Corbetta }
\medskip
{\footnotesize
\centerline{CASA- Centre for Analysis, Scientific computing and Applications,}
  \centerline{Department of Mathematics and Computer Science,
Eindhoven University of Technology,}
 \centerline{ P.O. Box 513, 5600 MB Eindhoven, The Netherlands,}
 \centerline{Department of Structural, Geotechnical and Building Engineering,} 
 \centerline{Politecnico di Torino, Corso Duca degli Abruzzi 24,
10126 Torino, Italy}   
} 

\medskip

\centerline{\scshape Adrian Muntean}
\medskip
{\footnotesize
 \centerline{CASA- Centre for Analysis, Scientific computing and Applications,}
  \centerline{ICMS - Institute for Complex Molecular Systems}
  \centerline{Department of Mathematics and Computer Science,
Eindhoven University of Technology,}
 \centerline{ P.O. Box 513, 5600 MB Eindhoven, The Netherlands,}

}

\medskip

\centerline{\scshape Federico Toschi}
\medskip
{\footnotesize
  \centerline{Department of Physics and Department of Mathematics and
    Computer Science,} 
\centerline{Eindhoven University of Technology, P.O. Box 513, 5600 MB Eindhoven, The Netherlands,}
  \centerline{CNR-IAC, Via dei Taurini 19, 00185 Rome, Italy,}
\medskip
}
\centerline{\scshape Kiamars Vafayi}
\medskip
{\footnotesize
  \centerline{CASA- Centre for Analysis, Scientific computing and Applications,}
  \centerline{Department of Mathematics and Computer Science,
Eindhoven University of Technology,}
 \centerline{ P.O. Box 513, 5600 MB Eindhoven, The Netherlands,}

}

\bigskip



\begin{abstract}
Focusing on a specific crowd dynamics situation, including real life
experiments and measurements, our paper targets a twofold aim: (1) we
present a Bayesian probabilistic method to estimate the value and the
uncertainty (in the form of a probability density function) of
parameters in crowd dynamic models from the experimental data; and (2)
we introduce a fitness measure for the models to classify a 
couple of model structures (forces) according to their fitness to the
experimental data, preparing the stage for a more general
model-selection and validation strategy inspired by probabilistic data
analysis. Finally, we review the essential aspects of our experimental
setup and measurement technique.
\end{abstract}

\newpage

\section{Introduction}
\subsection{Background}
Crowd models are powerful tools to explore the complex dynamics which characterizes the motion of pedestrians, cf.~e.g.~the overview~\cite{schadschneider2010stochastic} and references cited therein.  Understanding how crowds move and eventually being able to predict their behavior under given  (possibly extreme) conditions  becomes an increasingly important  matter for our society. Reliable mathematical models for crowds would be of great benefit, for instance, to increase pedestrian comfort (ensuring a regular flow motion at acceptable densities), security (evacuation assessments) and structural serviceability (crowd-structure interaction), in particular if the theoretical information/forecast is in real-time agreement with the actual crowd behavior. 

Aiming at quantitative models, a proper assessment of the uncertainty is needed given experimental data (e.g., in the form of crowd recordings) together with a model or a collection of models.   This paper treats a basic crowd scenario: pedestrians crossing an U-shaped landing (corridor) in a defined direction. In particular, the questions we  consider  here relate to a low density regime, in which pedestrians can be considered as moving alone, being just influenced by their desires as well as the geometry of the (built) environment in their surroundings (``rarefied gas'' regime).  
We explicitly wonder (in a quantitative sense specified afterwards):  do pedestrians \textit{interact} with walls? How does the presence of  walls affect the pedestrian motion? Are walls purely repulsive, or could they be attractive?  

To address these (and related) questions, we made many video recordings of the landing considered (see Appendix~\ref{ap}) and took as test models 7 different evolution equations describing the pedestrian-wall interaction.
\subsection{Simple crowd models. Ensembles}
Since the movement of  pedestrians is to a large extent
non-deterministic, any model that describes the detailed motion of
pedestrians requires the introduction of some elements of noise. To address this issue, one may choose to describe the dynamics from a ``coarse grained'' point of view, hence either at the mesoscopic scale or at the   macroscopic scale. For instance, when a macroscopic level of description is used, a balance equation for the density of pedestrians  (possibly supported by balance of momentum, see e.g.~\cite{Dafermos} for details on balance laws) models the evolution of the crowd, while the detailed microscopic behaviors are averaged out. 
In principle, the latter equations can be  derived directly from the microscopic equations, although a problem
remains: microscopic models for crowd dynamics (see, for instance, the
social force model proposed by Helbing and Molnar~\cite{helbing1995social}) 
describe the detailed motion of pedestrians, but to which extent the
microscopic details are relevant to capture the intrinsic crowd
patterns at larger scales and what is the role of noise and uncertainty? In other words, which details of the microscopic information are necessary to capture the  behavior of the crowd?

Usually, in an attempt to keep into account  the observed
non-deterministic behavior of the crowds, noise is added to 
deterministic models turning them into
Langevin-like equations for the so-called active Brownian particles (see
e.g.~\cite{Lutz,schadschneider2010stochastic}) or in measure-valued evolutions (see e.g.
\cite{banks2012least,evers2014well,bellomo2012modeling}). 

Since we cannot  describe deterministically the motion of pedestrians, we opt for a different strategy. We choose to construct a probabilistic ensemble of pedestrians (referred
here as the {\em ensemble}\footnote{By ``ensemble'' we understand a collection of the copies of a system distributed according to a probability distribution function (pdf).}) whose properties can be directly
deduced in a quantitative manner on the basis of experimental observations. 

We construct such crowd ensemble by the means of a connection to simple
evolution models, which can be cast in the form of a potential of
interaction of pedestrians with obstacles or amongst themselves
together via Newtonian dynamics. The potential or force in the model
is characterized by a set of parameters, whose statistical
properties can be obtained by comparing the model with well-controlled
experimental data. Thus the model and the probability distribution of
the model parameters (inferred from data)  define our {\em crowd
  ensemble}. In this way we cast {\em the force estimation problem}
into a well defined procedure of probabilistic data analysis, very
much inspired by~\cite{skilling1998probabilistic,sivia1996data}. A related approach for
traffic-flow models is mentioned in~\cite{hoogendoorn2010calibration} and references therein.

\subsection{Aim of the paper}
We focus our attention on the effects of walls (obstacles) on pedestrian
motion for a specific crowd dynamics scenario  (extensively described in Section~\ref{sect-exp-data} from a qualitative point of view, and in the Appendix~\ref{ap} from a technical point of view). Once the effect of walls is understood, considering pedestrian-pedestrian interactions in the same scenario is expected to be easier.

Note that even standard crowd dynamics situations are actually rather complex, in particular, the  following aspects among others need to be considered carefully:
\begin{itemize}
\item[(i)] the motion of pedestrian is complex and influenced by many sources, among others: desires/aims, interaction with geometry and interaction with neighboring peers;
\item[(ii)] the effects of these interaction appears simultaneously and in an entangled way.
\end{itemize}
To attempt to disclose the cause-effect relations in this complex motion, we choose a step by step approach; therefore, we opt to look exclusively at situations in which the interaction among  pedestrians is absent and the motion is fully regulated by own desires and neighboring geometry. As a clear consequence, this study sets a possible stage for the analysis of pedestrian-pedestrian interactions which is inevitably perturbed by the effects mentioned at (i) and (ii). To make the complete dynamics approachable, the hypothesis of linear superposition of effects has to be made.

In social-force models~\cite{helbing1995social,Degond_JSP,Mou, Sey1,BrunoCorbetta14P2} (for overviews on the matter see, e.g.,~\cite{Duives,schadschneider2010stochastic}), pedestrians move according to a Newtonian dynamics; in particular, in the absence of neighboring peers, the  force acting on a single particle can be expressed as:
\begin{equation}
F=F_{v_{d}}+F_{wall},
\end{equation}
where             
\begin{itemize}
\item $F_{v_{d}}$ is a force which keeps into account the desires of the pedestrians in terms of the direction he/she is willing to follow. Usually, this term induces a relaxation of the velocity $v$ of the pedestrian towards a background \textit{desired velocity} field $v_{d}=v_d(x,y)$. The desired velocity field usually drives the pedestrian all over the domain toward a given ``desired'' target. In formulas, this term reads as $F_{v_{d}}=(v-v_{d})/\tau$, where  $\tau$ is a characteristic relaxation time.
\item $F_{wall}$ describes the interactions pedestrians have with walls. This term  doesn't just model the impenetrability of the latter, rather it is aimed at taking into account  the will of pedestrians to maintain a certain distance from walls. 
\end{itemize}

Structure and parameters dependence of $v_d$ and $F_{wall}$ shall be assumed. In the following, we suppose $v_d$ to be parametrized by its magnitude (the \textit{desired speed} $|v_d| \equiv \mbox{\textit{const}}$), whilst its direction is kept as a model feature.

On the other hand, $F_{wall}$ is assumed to be a  sum of forces pointing outwards every wall  in the particle  proximity, i.e.
\begin{equation*}
F_{wall} = \sum_{w \in \mbox{Walls}}F_{wall}^w.
\end{equation*}
One usually supposes  that every contribution $F_{wall}^w$  has a fixed functional form which depends on the distance between the pedestrian and the wall, and is parametrized by a set of $N_{p}$ parameters $\vec P=\left\{ P_{i}\right\} $. Note, however, that different, more general, forms of $F_{wall}$ can be chosen; see e.g.~\cite{jaklin2013real}. 

In the present work, our main purpose is to  estimate the probability distribution functions of the model parameters $v_{d}$ and $\left\{
P_{i}\right\} $ from experimental data (cf. Section
\ref{sect-exp-data} and Appendix~\ref{ap}).

\subsection{Experimental data}\label{sect-exp-data} 
\begin{figure}
\begin{raggedright}
\includegraphics[width=6cm]{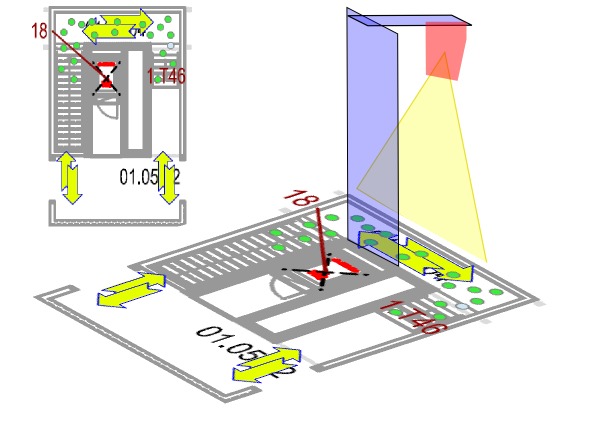}\includegraphics[width=7cm]{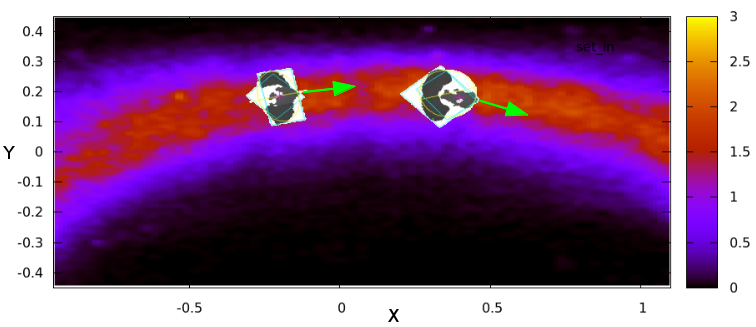}
\par\end{raggedright}

\caption{Left: Representation of the landing map. Right:  Actual view from the camera.  Here two pedestrians are
  detected as they walk through the landing.  Multi-pedestrian events,
  such as the one shown here for illustration, are actually filtered
  out, so that only trajectories involving only one
  pedestrian at a time are considered. The density plot in the background shows the
  density count of pedestrian, as measured from our camera and
  distributed over the 2D space of the
  landing, after a long time.\label{fig:The-view-from}}

\end{figure}

The kinematic data  referring to trajectories of pedestrians (positions, velocities, accelerations) walking through a rectangular landing   (see Figure
\ref{fig:The-view-from}. Consider Appendix
\ref{ap} for details), are acquired via an over head recording  camera. Due to the geometry of the setting, trajectories of pedestrians tend to  bend slightly as the landing is crossed. This is a consequence of the  presence of $90$ degree turns at the entrance and at the exit of the landing.

In the following, since we focus our attention on the walls-pedestrian interaction, we consider only data concerning a single pedestrian (i.e.~appearing alone in the camera field of view) and going in a specific direction (specifically, from left to the right) are considered. In other words,  all data referring to situations in which more than one pedestrian are present at a time are not taken into consideration.

The final measured data consists of a set of $N_{T}$ pedestrian trajectories ($N_T\approx 1500$)
\begin{equation*}
\left\{ T_{i}|\:1\leq i\leq N_{T}\right\}.
\end{equation*}
Every trajectory 
is  a set of recorded points in the $2+1$-dimensional space-time of the landing (i.e., a given pedestrian having trajectory $T_i$ has spatial position $(x_{k}^{i},y_{k}^{i})$ at time instant $t_{k}^{i}$). In formulas, we describe a trajectory like
\begin{equation}\label{eq-trajectory-single}
T_{i}=\{(x_{k}^{i},y_{k}^{i},t_{k}^{i})|\:1\leq k\leq M_{i}\},
\end{equation}
where $M_{i}$ is the number of points in the considered trajectory.

\begin{remark}\label{Rem}
Looking back at the considered model parameters  (i.e., the desired speed $|v_{d}|$ and the wall force parameters, $\left\{
P_{i}\right\} $), we observe that $|v_{d}|$ is a parameter specific of the
trajectories, i.e.~to every trajectory it corresponds a specific value $|v_{d}|$ that needs
to be estimated. In contrast, one can think of $\left\{P_{i}\right\} $
as a set of global parameters, in the sense that they are shared
between all trajectories. This would suggest the use of a two-steps optimization procedure; see Section~\ref{2steps}. In this paper, we decide to treat all parameters in the same way, and thus we postpone the two-steps optimization idea for a later approach. 
\end{remark}

\subsection{Structure of the paper} 
The paper is organized as follows: Section~\ref{prob-method} contains
the working methodology which is based on probability estimates guided by Bayes theorem; in Section~\ref{selection} we point out how Bayes theorem can be used for model selection. The main result of this paper is presented in Section~\ref{result_forces}: following the described working methodology, parameters of a number of simple wall force models for our crowd scenario are estimated. Moreover, on this basis, models are compared quantitatively and qualitatively. 
Section~\ref{sect-discussion} discusses  the obtained results. Finally, the measurement technique and the collected experimental data are described in Appendix~\ref{ap}.

\section{Probabilistic data analysis of crowd ensembles}\label{prob-method}
In this section, we introduce the methodology behind the probabilistic data analysis used here. For more details, we refer the reader to the introductory guide by Skilling~\cite{skilling1998probabilistic} or to the mathematical background presented by Sivia in~\cite{sivia1996data}. 
 
\subsection{Probability estimates. Bayes theorem}
We denote all the measured data by $D$ and all prior information by
$I$. In other words, while $D$ encloses all the information acquired in the measurement process, $I$ includes the assumptions made on model and thus on the type of dynamics. 

Our goal is to identify the parameters for a considered set of pedestrian models ($|v_{d}|,\left\{P_{i}\right\}$) for the considered scenario. This task can be performed by  estimating the \textit{posterior} probability law
\begin{equation*}
Prob\left(v_{d},\left\{ P_{i}\right\}|D,I\right),
\end{equation*} 
which describes the probability associated to the parameters of a considered model being conditioned on data $D$ and all prior information $I$. Such probability law can be either peaked around a given maximum value, which does correspond to the solution of the estimation problem, or can be dispersed. In such case the mean value and the standard deviation of the law can be used as a fair representation of the full distribution. 

By means of Bayes theorem (see e.g.~\cite{cox1961algebra,citeulike:3446780,jaynes2003probability}), the posterior probability can be related to other probabilities of easier computation (or even known already). The theorem reads as
\begin{equation}\label{Bayes}
Prob\left(v_{d},\left\{ P_{i}\right\} |D,I\right)=\frac{Prob\left(D|v_{d},\left\{ P_{i}\right\} ,I\right)Prob\left(v_{d},\left\{ P_{i}\right\} |I\right)}{Prob\left(D|I\right)}.
\end{equation}
The probabilities involved in the l.h.s~are respectively
\begin{itemize}
\item the \textit{likelihood function} $Prob\left(D|v_{d},\left\{ P_{i}\right\} ,I\right)$;
\item the \textit{prior probability} $Prob\left(v_{d},\left\{ P_{i}\right\} |I\right)$; 
\item the \textit{data evidence} $Prob\left(D|I\right)$. 
\end{itemize}
In the following subsections such probabilities are extensively used and details regarding their computations are given. It is worth to notice that, since the data $D$ is assumed as given, the data evidence plays the role of a mere normalization constant and, as a consequence, we can write
\begin{equation}
Prob\left(v_{d},\left\{ P_{i}\right\} |D,I\right)\thickapprox Prob\left(D|v_{d},\left\{ P_{i}\right\} ,I\right)Prob\left(v_{d},\left\{ P_{i}\right\} |I\right),
\end{equation}
to stress the fact that the quantities playing a significant role in parameter estimation are just the likelihood function and the prior probability.

\subsection{Likelihood function. Misfit norm(s)}
The likelihood function  measures how well the model with the given parameters fits the
data. We distinguish between four different schemes to compare data
and models:
\begin{enumerate}
\item[(i)] positions in trajectory data versus positions predicted in
  the model;
\item[(ii)] velocities obtained from data versus 
  velocities as calculated from the model;
\item[(iii)] acceleration deducted from data versus acceleration in
  the model;
\item[(iv)] a combination of the previous three.
\end{enumerate}
It is worth to remark that the third scheme has the advantage of not requiring the computation of  the full trajectories generated by the model. On the other hand,
acceleration data are usually more noisy, as a consequence of the double time
differentiation of pedestrian trajectories (which is never exempt from noise).

The likelihood function can be obtained from the Principle of Maximum Entropy (MaxEnt) 
once  the kind of noise in the data is assumed, see \cite{sivia1996data} for details. In particular, assuming Gaussian noise, the likelihood function results in
\begin{equation}\label{like}
Prob\left(D|v_{d},\left\{ P_{i}\right\}
,I\right)=\Pi_{k=1}^{N_{k}}\left(\sigma_{k}\sqrt{2\pi}\right)^{-1}\exp\left(\frac{-\sum_{k=1}^{N_{k}}\left(d_{k}-m_{k}\right)^{2}}{2\sigma_{k}^{2}}\right),
\end{equation}
where $d_{k}$ is the acceleration in the trajectory data, at sample
$k$, and $m_{k}$ is the acceleration provided by the model with parameters $|v_{d}|$ and $\left\{ P_{i}\right\} $ at the same point.
According to the adopted notation, $\sigma_{k}$ is the error estimation, or standard deviation, for the experimental acceleration
$d_{k}$.

We consider the likelihood function  for two different
assumptions on the noise in the data: (i) Gaussian noise and (ii)
Exponential noise. Different assumptions on the noise may be made\footnote{Note that in the Bayesian framework the noise is part of the model.}.
The choices (i) and (ii) seem to be more common in literature.  In this
paper, we decide to employ (i) and to leave for further investigations
the structure of the noise in our data (see~\cite{ours2014}).

\subsubsection{Gaussian noise}

The misfit function $\chi^2$ is defined such that

\[
\chi^{2}=-2\log\: Prob\left(D|v_{d},\left\{ P_{i}\right\}
,I\right)+C(\sigma_{k},N_{k})
\]
or 

\[
Prob\left(D|v_{d},\left\{ P_{i}\right\}
,I\right)\thickapprox\exp\left(\frac{-\chi^{2}}{2}\right).
\]

Therefore, for the case of Gaussian noise in the data, we have the
$\ell_{2}$ norm for the ``distance'' between the model and data;

\[
\chi^{2}=\frac{\sum_{k=1}^{N_{k}}\left(d_{k}-m_{k}\right)^{2}}{\sigma_{k}^{2}}.
\]

Since the logarithm is a monotonically increasing function, finding
the maximum of $Prob\left(D|v_{d},\left\{ P_{i}\right\} ,I\right)$ is
equivalent to finding a minimum for $\chi^{2}$. It becomes more
simplified if $\sigma_{k}$ are equal to $\sigma$,

\[
\chi^{2}=\frac{1}{\sigma^{2}}\sum_{k=1}^{N_{k}}\left(d_{k}-m_{k}\right)^{2}
\]
or more reasonably since we would intuitively expect that the error
estimate be proportional to the data
$\sigma_{k}=\sigma\left|d_{k}\right|$ and define a new misfit norm by
dividing the previous one by $N_{k}$;

\[
\chi^{2}=\frac{1}{N_{k}\sigma^{2}}\sum_{k=1}^{N_{k}}\left(\frac{d_{k}-m_{k}}{d_{k}}\right)^{2}.
\]

\begin{remark} Since  we do not study explicitly the absolute magnitude of the noise, we take everywhere in the paper $\sigma$ to be  $1$.
\end{remark}
This form of $\chi^{2}$ has the interesting property that if
$m_{k}=d_{k}(1+\epsilon\sigma)$, i.e., if the model misses the data by
a (small) fraction of $\epsilon$ of the error estimate, then we have

\[
\chi^{2}(\epsilon)=\epsilon^{2}.
\]

\subsubsection{Exponential noise}

The exponential noise in the data corresponds to

\[
Prob\left(D|v_{d},\left\{ P_{i}\right\}
,I\right)=\Pi_{k=1}^{N_{k}}\left(2\sigma_{k}\right)^{-1}\exp\left(-\sum_{k=1}^{N_{k}}\frac{\left|d_{k}-m_{k}\right|}{\sigma_{k}}\right),
\]
therefore the misfit $\chi^{2}$ in this case will be the $l_{1}$ norm

\[
\chi^{2}=\sum_{k=1}^{N_{k}}\frac{\left|d_{k}-m_{k}\right|}{\sigma_{k}}.
\]
Again if we assume $\sigma_{k}=\sigma\left|d_{k}\right|$ and divide by
$N_{k}$ we obtain

\[
\chi^{2}=\frac{1}{N_{k}\sigma}\sum_{k=1}^{N_{k}}\left|\frac{d_{k}-m_{k}}{d_{k}}\right|.
\]
A similar calculation as the one we did for Gaussian noise for the
deviation $m_{k}=d_{k}(1+\epsilon\sigma)$, yields
\[
\chi^{2}(\epsilon)=\epsilon.
\]

The exponential distribution is less centrally distributed than a
Gaussian. Consequently, the $\ell_{1}$ norm is more robust and it is
expected to fit better data that contains a number of ``wildly''
distributed points.

With our choice of the misfit functions $\chi^{2}$, we are {\em de
  facto} pushing forward an empirical probability measure (defined by
the data), from the parameter space onto the real line.  This allows us
to compare in a natural fashion different models.

\subsection{Prior probability}
\subsubsection{Backgorund} 
The prior probability $Prob\left(v_{d},\left\{ P_{i}\right\}
|I\right)$ encodes our prior state of knowledge on the parameters,
before taking into account the acquired data $D$. For most practical purposes
we can suppose that it is a constant over the range of parameters that
we consider acceptable.  More precisely we can use the symmetry group
transformations and/or the principle of maximum entropy to assign the
prior probabilities.  From translation symmetry arguments, it turns
out that for position parameters, like coordinates, a uniform prior
distribution over the expected range is the optimal choice. For scale
parameters, the right prior is the Jeffreys distribution
$Prob\left(p\right)\propto\frac{1}{p}$, which is a consequence of
scale invariance symmetry~\cite{jaynes2003probability,jaynes1968prior,jaynes1973well}.  As long as the
data is of good quality and the range of parameters is chosen well,
one expects that the effect of the likelihood function dominates, and
the posterior probability is less sensitive to the exact choice of the
prior probability.

\subsubsection{Law of large numbers}\label{large}

Since the number of trajectories measured can be arbitrarily large, we
expect that the law of large numbers applies and that one can use an
optimization procedure on each single trajectories separately and then
employ the resulting histogram to estimated the parameters as the
probability distribution for the value of the parameters. Thus having access to a large number of trajectories simplifies the estimation scheme.

\subsection{Two-steps optimization}\label{2steps}

As mentioned in Remark~\ref{Rem}, the experimental technique and setup indicate  that there is an
intrinsic difference between the parameters $v_{d}$ and $\left\{
P_{i}\right\}$, the former being trajectory dependent and the later
being global. One way to take this into account is to estimate firstly the
parameters for each trajectory, in particular $v_{d}$, obtaining
$v_{d,i}$
As the second step, one can afterwards use the data $\left\{
\left(T_{i},v_{d,i}\right)|\:1\leq i\leq N_{T}\right\} $ for globally
estimating $\left\{ P_{i}\right\} $ and calculating the global fitness
of the model to data. We do not follow this approach here, rather we treat all parameters equally.

\subsection{Simulated annealing}

The parameter space is in general multidimensional, and posterior
probability is likely to be multi-modal, where the probability
maximums (or the minimum of the misfit norm) are generally not
analytically solvable. One can use the simulated annealing method~\cite{kirkpatrick1983optimization} to tackle this problem; in analogy with thermodynamics,
one supposes that the misfit norm is an energy function and uses the
Metropolis Monte Carlo algorithm~\cite{metropolis1953equation} to sample the points in
parameter space according to the Boltzmann-Gibbs distribution at a
given temperature, $T$:

\[
Prob_{T}\left(v_{d},\left\{ P_{i}\right\} \right)\propto e^{\frac{-\chi^{2}}{T}}.
\]

Clearly by setting $T=1$ we get the same distribution as the desired
likelihood probability. The basic idea of simulated annealing is to
start with $T>1$ and then reduce the temperature dynamically until
$T=1$. In this way, starting from a more ``energetic'' point, it is more
likely to overcome the local minimum traps of the misfit function,
and reach the most significant parts around the global
minimum. Afterwards we will continue the sampling with $T=1$ to obtain
the points in parameter space distributed according to the posterior.
In order to achieve good convergence, it might be necessary to repeat
the whole annealing procedure several times and by starting from
different random initial points. The resulting distribution can be
used to obtain the marginal probability distribution of for example
$v_{d}$;

\[
Prob\left(v_{d}\right)=\int Prob_{T=1}\left(v_{d},\left\{
P_{i}\right\} \right)\:\prod dP_{i},
\]
which can be directly plotted or used to calculate the average and the
standard deviation for the parameter;

\[
\overline{v_{d}}=\left\langle v_{d}\right\rangle =\int v_{d}\,
Prob_{T=1}\left(v_{d},\left\{ P_{i}\right\} \right)\:\prod dP_{i}
\]

\[
\sigma_{v_{d}}^{2}:=\left\langle
\left(v_{d}-\overline{v_{d}}\right)^{2}\right\rangle =\int v_{d}^{2}\,
Prob_{T}\left(v_{d},\left\{ P_{i}\right\} \right)\:\prod
dP_{i}\,-\,\overline{v_{d}}^{2}.
\]

In the framework of this paper, using many trajectories as data, we perform the per-trajectory optimization approach mentioned in Section~\ref{large} and we thus do not require the simulated annealing approach.

\section{Selection and hierarchy of models}\label{selection}

\subsection{Background}

Suppose we have two models $M_{A}$ and $M_{B}$.  In such case, we can apply the
Bayes theorem to obtain

\[
Prob\left(M_{A}|D,I\right)=\frac{Prob\left(D|M_{A},I\right)Prob\left(M_{A}|I\right)}{Prob\left(D|I\right)}
\]

and, similarly for $M_{B}$. Therefore, we have 

\[
\frac{Prob\left(M_{A}|D,I\right)}{Prob\left(M_{B}|D,I\right)}=\frac{Prob\left(D|M_{A},I\right)Prob\left(M_{A}|I\right)}{Prob\left(D|M_{B},I\right)Prob\left(M_{B}|I\right)}.
\]

In general, the two models will have different set of parameters,
$\overrightarrow{P}_{A}$ and $\overrightarrow{P}_{B}$, respectively. We
have for the data $D$ and assuming model $M_{A}$ that 

\[
Prob\left(D|M_{A},I\right)=\int\,
Prob\left(D|\overrightarrow{P}_{A},M_{A},I\right)\,
Prob\left(\overrightarrow{P}_{A}|M_{A},I\right)\,
d\overrightarrow{P}_{A},
\]

hence

\[
\frac{Prob\left(M_{A}|D,I\right)}{Prob\left(M_{B}|D,I\right)}=\frac{Prob\left(M_{A}|I\right)}{Prob\left(M_{B}|I\right)}\times\frac{\int\,
  Prob\left(D|\overrightarrow{P}_{A},M_{A},I\right)\,
  Prob\left(\overrightarrow{P}_{A}|M_{A},I\right)\,
  d\overrightarrow{P}_{A}}{\int\,
  Prob\left(D|\overrightarrow{P}_{B},M_{B},I\right)\,
  Prob\left(\overrightarrow{P}_{B}|M_{B},I\right)\,
  d\overrightarrow{P}_{B}}.
\]

$Prob\left(M|I\right)$ indicates our prior probability for the model
$M$.  The integrals over the likelihood function can be calculated
with the same stochastic procedure as explained in the section for
simulated annealing. We see now that the parameters priors
$Prob\left(\overrightarrow{P}|M,I\right)$ play a role, therefore it is
important to make sure they are assigned properly, especially when the
two models are nearly identical. We take here
$\frac{Prob\left(M_A|I\right)}{Prob\left(M_B|I\right)}=1$ but, in
principle, this ratio does not need to be one and it can therefore be used for
the updating procedures provided previous information are available.

\subsection{Law of large numbers}

As in the previous section, we use again the fact that we are in a
law of large numbers regime and, after the optimization procedure on
each single trajectory separately, we use the resulting histogram of
the minimum value of the misfit function as the criteria for the model
selection. If only one number is required to compare the models, it
will be then the average of
\begin{equation}\label{L}
L=\chi^{2}
\end{equation} 
in such histograms, for each model, the smaller values are an
indication that the model fits better to the data.

\begin{figure}[t]
\begin{centering}
\includegraphics[width=8cm]{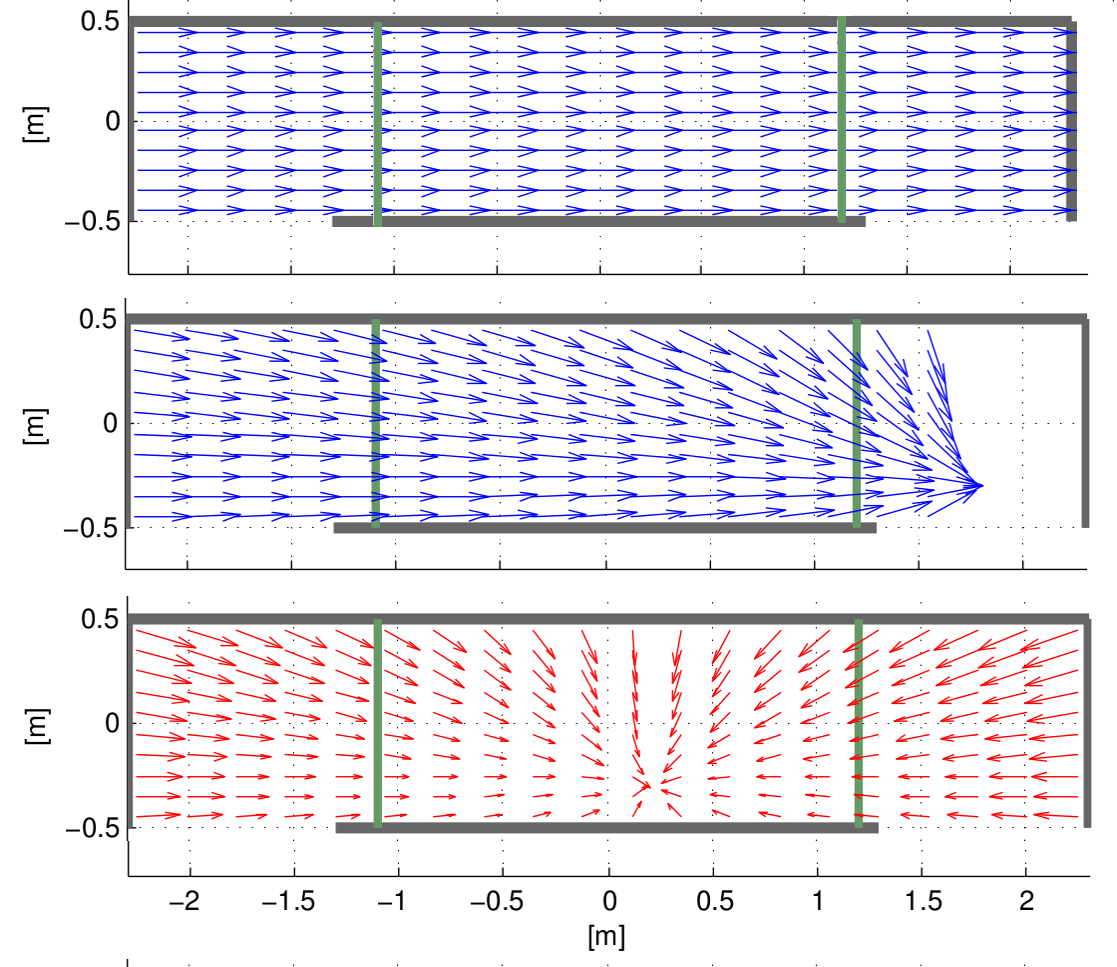}
\par\end{centering}

\caption{The setup of the models as seen from the camera. From up to
  bottom; the desired velocity field of $M_{1}$, the desired velocity
  field of $M_{3}$ and the force field of
  $M_{5}$.\label{fig:The-setup-of}}
\end{figure}

\section{Estimating wall forces}\label{result_forces}

\subsection{Choice of models}

We consider $7$ different \textit{simple}  crowd models which we denote by the symbols $M_{1},M_{2},\ldots,M_{7}$, respectively. We apply the probabilistic parameter estimation technique described in Section~\ref{prob-method} - Section~\ref{selection}, together with our experimental test scenario described in Appendix~\ref{ap}. 
 
The chosen models aim at reproducing the basic aspects of the pedestrian motion for individuals going from the left hand side to the right hand side of the landing. As previously mentioned (see also Figure~\ref{fig:The-view-from} (right)), pedestrian trajectories are slightly bending as a consequence of the landing shape. 

$M_1-M_7$ feature different levels of complexity as more and more phenomenological aspects are introduced. In the simplest case, a rightward directed homogeneous velocity field is considered. 
Hence, terms to take into account the wall forces and also to keep track of the shape of the corridor (e.g. via the desired velocity) are added.

More precisely, the models $M_1-M_7$ are: 
\begin{itemize}
\item [$M_{1}$:] An homogeneous desired velocity field parallel to the span-wise walls is adopted. The repulsion force of the walls is neglected. The relaxation time is hereby adopted as fixed and, according to~\cite{helbing1995social}, the value  $\tau=0.5s$ is chosen. As a consequence, the parameter to estimate is the desired speed $|v_d|$, i.e.~$\overrightarrow{P}_1 = \{\}$; see Figure~\ref{fig:The-setup-of} (top).
\item [$M_2$:] An homogeneous velocity field analogous to the one in $M_{1}$ is chosen, however the relaxation time $\tau$ is also treated as a parameter to be estimated, i.e.~$\overrightarrow{P}_2 = \{\tau\}$. 
\item [$M_{3,4}:$] These models are similar - respectively - to $M_{1}$ and $M_{2}$, and they differ from the latter as the  desired velocity
  field is such that at every point the desired velocity vector is
  directed towards the exit point, i.e.~$\overrightarrow{P}_3 = \{\}$ and $\overrightarrow{P}_4 = \{\tau\}$; see Figure~\ref{fig:The-setup-of} (middle).
\item [$M_{5}$:] This model features exclusively the wall force $F_{wall}$ and no relaxation toward a desired velocity field. This force is directed radially in any point toward the center $C$ of the ``lower'' wall.

More precisely, the force at a point $(x,y)$ is given by
\[
F_{wall}(x,y)=m(x,y)\vec\iota(x,y),
\]
where $m(x,y)$ is the magnitude of the force and $\vec\iota(x,y)$ is a unit vector field pointing towards $C$. Moreover, we assume
that
\[
m(x,y)=A\sum_{i=1}^3e^{-kd(r_i(x,y))}+B,
\]
where $A,B$ and $k$ are parameters to be estimated and $d(r_i(x,y))$ is the distance of the point $(x,y)$ to the center of the lower wall, i.e.~$\overrightarrow{P}_5 = \{A,B,k\}$. 
\item [$M_{6}$:] This is a particular case of $M_{5},$ where we suppose that
  the force magnitude is constant, i.e.
\[
m(x,y)=A.
\]
Here $A$ is the parameter to be estimated, in other words, $\overrightarrow{P}_6 = \{A\}$.
\item [$M_{7}$:] This is similar to $M_{5},$ but, in addition to the force, a  desired velocity is also taken into account. This desired velocity  field is chosen to be similar to one in the models $M_{3}$ and $M_4$,  pointing  towards the exit. As in $M_3$, $\tau=0.5$s, i.e.~$\overrightarrow{P}_7 = \{A,B,k\}$.
\end{itemize}

\subsection{Optimization method}

We use the $\ell_2$ norm for the misfit function (corresponding to the choice of Gaussian noise) and  a ``global'' 
brute-force-based optimization procedure to find the best fitting
parameters per trajectory.
{\em A priori} knowledge on the parameters range is
assumed and this defines a box in the parameters space. The box has
been evenly sampled and the cost function has been evaluated. A
gradient descent approach is then applied on the sample of minimum cost 
used to refine the result.

\begin{figure}
\includegraphics[width=10cm]{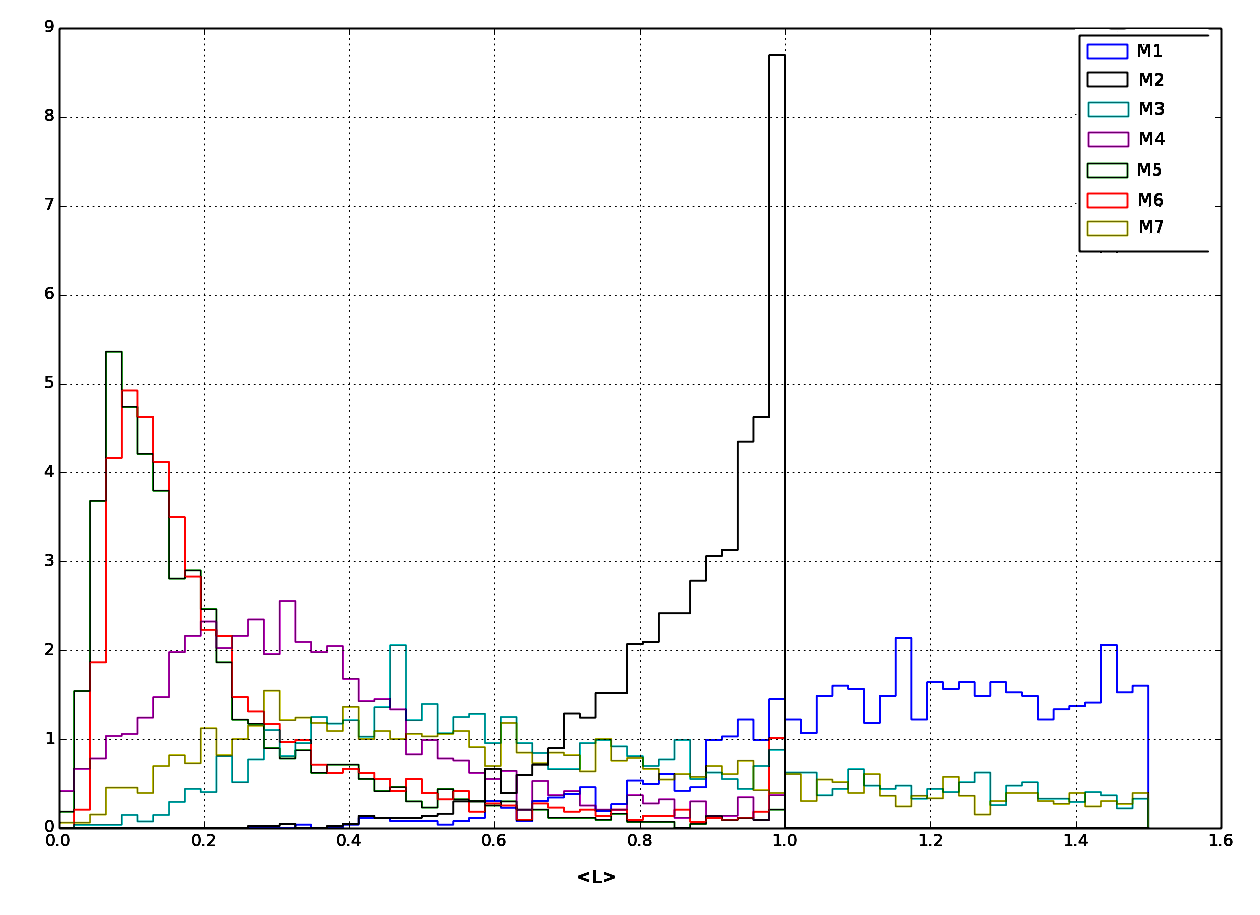}

\caption{Comparison of the models based on the misfit histogram
  distributions [counts (frequency) {\em vs.} misfit $L$ 
    cf. (\ref{L})].\label{fig:Comparison}}
\end{figure}

\begin{figure}
\includegraphics[width=11cm]{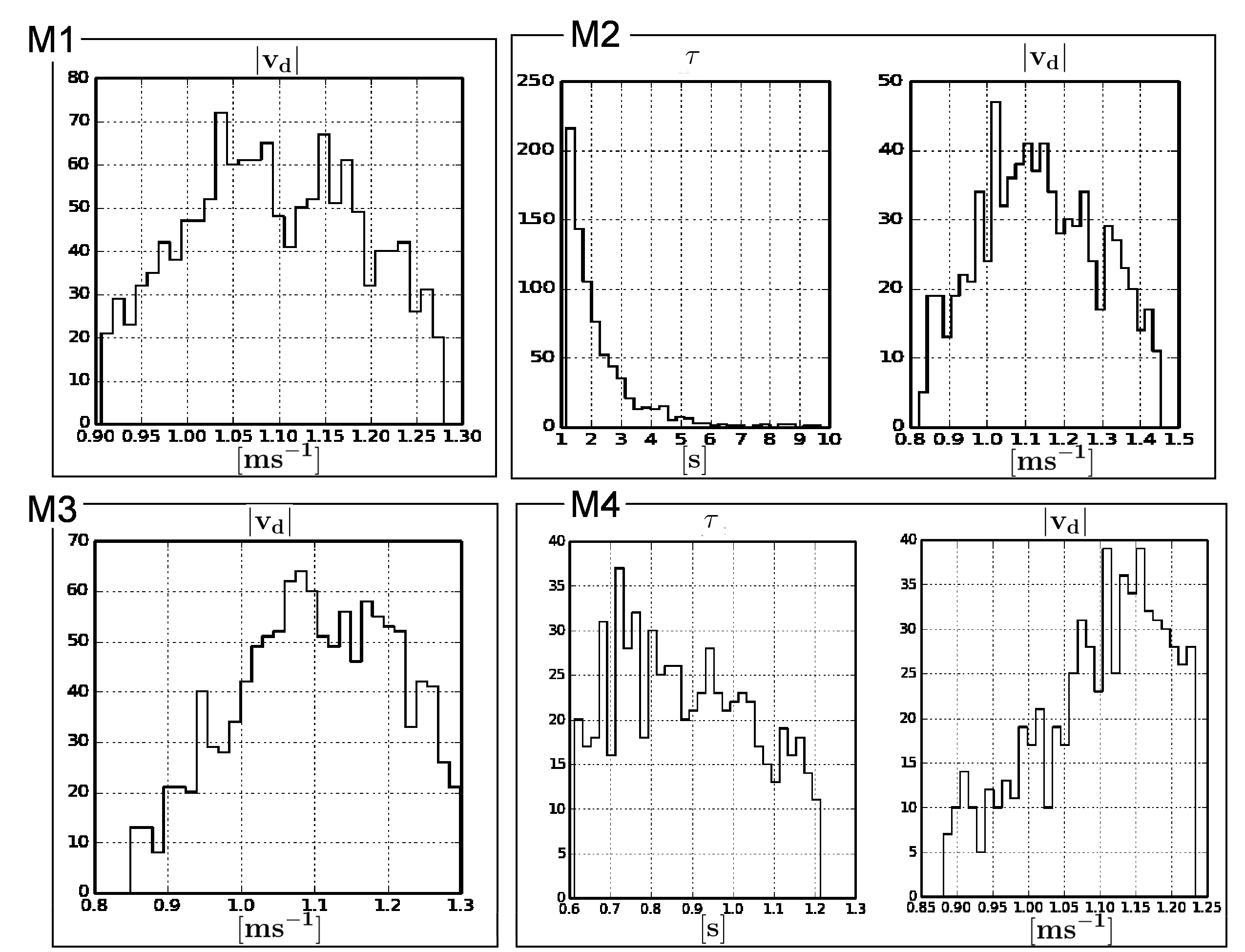}

\caption{Histogram distributions (counts (frequency) {\em vs.} values)
  for the value of parameters corresponding to models
  $M_1,\ldots,M_4$.\label{fig:Comparison-1}}
\end{figure}

\begin{figure}
\includegraphics[width=11cm]{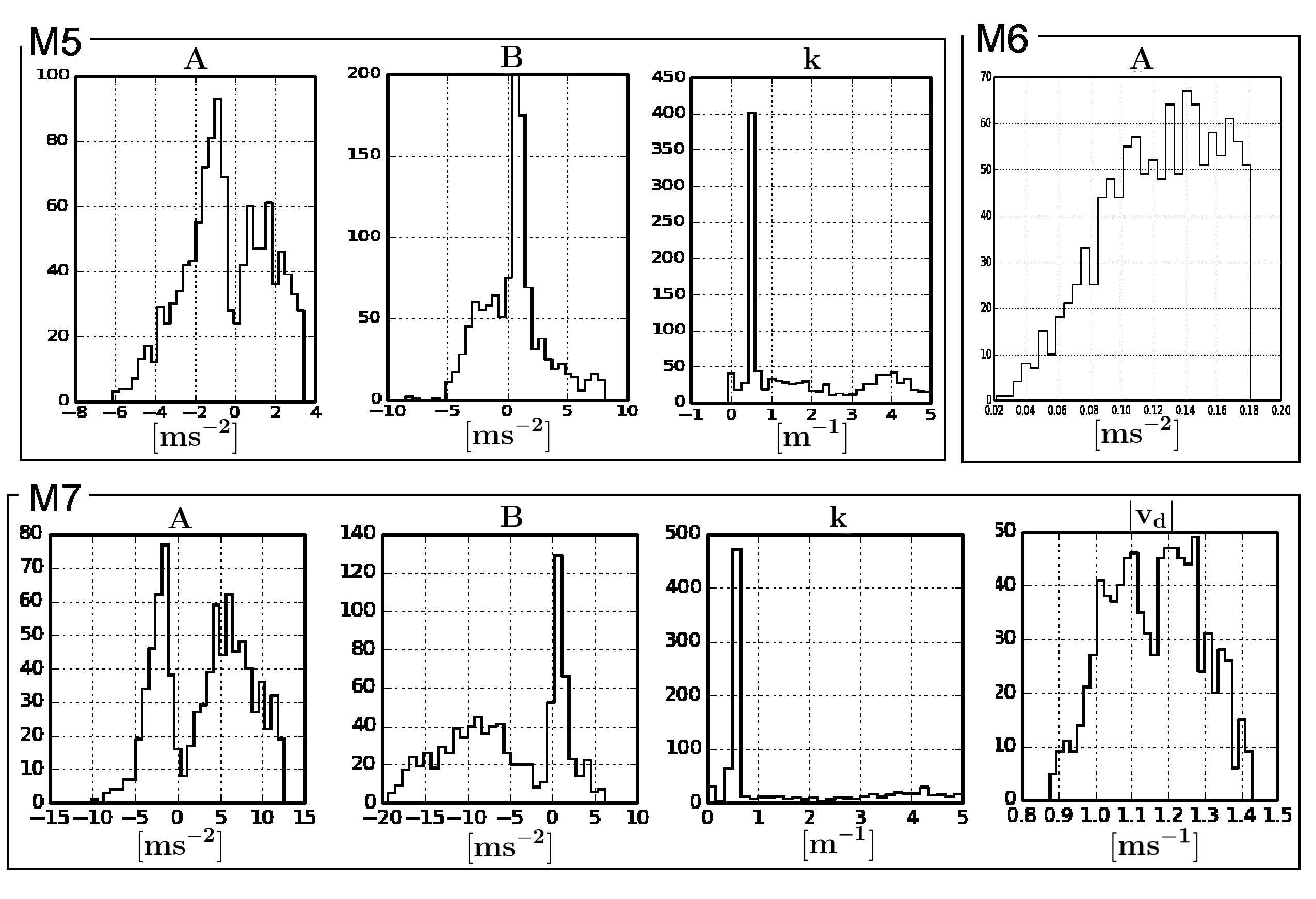}

\caption{Histogram distributions (counts (frequency) {\em vs.} values)
  for the value of parameters corresponding to models
  $M_5,\ldots,M_7$.\label{fig:Comparison-1-1}}
\end{figure}

In Figure~\ref{fig:Comparison}, we compare the models based on the
histogram distribution of the misfit function. In Figures~\ref{fig:Comparison-1} and~\ref{fig:Comparison-1-1}, empirical distributions of the value of the parameters are histogrammed. In Table~\ref{tab:Average-value-of} the average
value of the parameters and the misfit norms calculated from the
histograms are presented.

\newcommand{\avL}{\langle L\rangle}

\subsection{First observations}

Looking at the results of the estimation method, shown in Figure  \ref{fig:Comparison}, we observe that in
the sense of accordance with the data\footnote{As mentioned before, the property of having the least amount of
$\avL$, or, in other words; the more the distribution of $L$ gravitates towards $0$, the better the fit.} the two best performing models are the force only models, that is 
$M_{5}$ and $M_{6}$, then  followed by $M_{3}$.

The wide reduction in terms of $\avL$ in $M_{4}$ compared to $M_{2}$ shows the
importance of having a good\footnote{In this context, by ``good'' we mean small misfit or large likelihood.} desired velocity field, and that more complicated fields might reduce $\avL$ even further. However, it is still more than $\avL$ for $M_{5}$, even though $M_{5}$ uses a quite simple force field.

Comparing the outcome for models $M_{1}$ and $M_{2}$ (or similarly for
models $M_{3}$ and $M_{4}$), we notice that introducing the extra
parameter for the relaxation time $\tau$ decreases $\avL$, thus making a
better fit to the data.  However it has a negligible effect on the
magnitude of the desired velocity. The average $\tau$, on the other
hand, is rather dependent on the desired velocity field, as it can be
seen from a comparison of models $M_{2}$ and $M_{4}$, but again the
average magnitude of the desired velocity is not very sensitive to the
choice of the field and in fact varies very little across the models
$M_{1}$, $M_{2}$, $M_{3}$, $M_{4}$ and $M_{7}$. The estimated desired speed shows agreeing values  (see the histograms of $|v_d|$ in Figures~\ref{fig:Comparison-1} and~\ref{fig:Comparison-1-1}, and its average values in Table~\ref{Raz}) which, furthermore, are slightly overestimating the actual pedestrian velocities measured ($\langle v \rangle \approx 1.05 m/s$) consistent with the fact that they define a comfortable target velocity the pedestrians aim at achieving.

We found considerable correlations between the values of $A$ and $B$
in the models $M_{5}$ and $M_{7}$. This indicates that the data that
we have is not able to estimate these two parameters separately very
well, but their sum $A+B$ can be rather well estimated, which is
approximately (given that $k$ in $M_{5}$ and $M_{7}$ is small) what is
being done in $M_{6}$, where the histogram for $A$ is more sharply
peaked than the histogram for either $A$ or $B$ in the models $M_{5}$
and $M_{7}$.

\begin{table}
\begin{centering}
\begin{tabular}{|c|c|c|c|c|c|c|}
\hline 
Model & $A [ms^{-2}]$ & $B [ms^{-2}]$ & $k [m^{-1}]$ & $|v_d| [ms^{-1}]$ & $\tau [s]$ & $\avL [1]$\tabularnewline
\hline 
\hline 
$M_1$ & - & - & - & $1.11$ & - & $1.30$\tabularnewline
\hline 
$M_2$ & - & - & - & $1.18$ & $2.14$ & $0.83$\tabularnewline
\hline 
$M_3$ & - & - & - & $1.13$ & - & $0.81$\tabularnewline
\hline 
$M_4$ & - & - & - & $1.20$ & $1.40$ & $0.33$\tabularnewline
\hline 
$M_5$ & $-0.016$ & $0.55$ & $1.31$ & - & - & $0.20$\tabularnewline
\hline 
$M_6$ & $0.13$ & - & - & - & - & $0.25$\tabularnewline
\hline 
$M_7$ & $3.05$ & $-3.85$ & $1.02$ & $1.19$ & - & $0.73$\tabularnewline
\hline 
\end{tabular}\label{Raz}
\par\end{centering}

\caption{Average value of models parameters and the misfit
  norm\label{tab:Average-value-of}}
\end{table}

\section{Discussion}\label{sect-discussion}

Focusing on a specific pedestrian dynamics scenario, we presented
a procedure for estimating models parameters for pedestrian movement
based on cameras real life measurements. Basing the method on its
Bayesian foundations, we indicated how having access to a
large number of measurements simplifies and improves the parameter estimation and the models selection
procedures.

The desired velocity and external force acting on pedestrians are two
different modeling routes and ingredients, used sometimes separately,
and sometimes in combinations. For instance, in the social force model
both ingredients are usually present. In the simple landing setup
studied in this paper, we found that in case that we use simple models
based on force or desired velocity fields, the force-only models do a
significantly better job to match the data. The force-only models
apparently outperform also models that combine force and desired
velocities. This is presumably due to the fact that the slight
increased complexity of the model is not necessarily producing better fit to the data in general. The outcome of such increase in models complexity is nontrivial.

Possible extensions of this work can go in multiple directions: for
instance, one definitely needs to study other geometries and
experimental setups as well as  more detailed models allowing for instance the
interplay between pedestrian(s)-wall vs.~pedestrian-pedestrian
interaction. Last but not least, we expect that from the experimental
measurements much can be learned on the time and space structure of
the correlations in the pedestrian flow (to be followed by us in
 ~\cite{ours2014}).

Finally, it  worths noting that the approach presented here is not
exclusively meant for crowd dynamics applications. The parameter
identification procedure can be exploited in a large variety of
settings ranging from the tracking of cells motion in biological
flows, the motion of colloidal suspensions, the detection of
localization patterns of stress-driven defects in materials.

\appendix
\section{Experimental setup}\label{ap}
We provide hereby fundamental  information on the experimental setup and data used in this paper.

\subsection{Lagrangian measurement of pedestrian dynamics}

The experimental data considered in this paper, which have been used
as a reference to tune and to compare pedestrian models, have been
collected during a months-long experimental campaign. 

Heavily trafficked landing (see Figure~\ref{expsetup}), which connects the canteen to the dining area of the MetaForum building at 
Eindhoven University of Technology, has been recorded on 
full-time (24/7) basis. These recordings allowed us to gather the   statistics concerning pedestrian trajectories that we considered throughout this manuscript.

\putfig{.9}{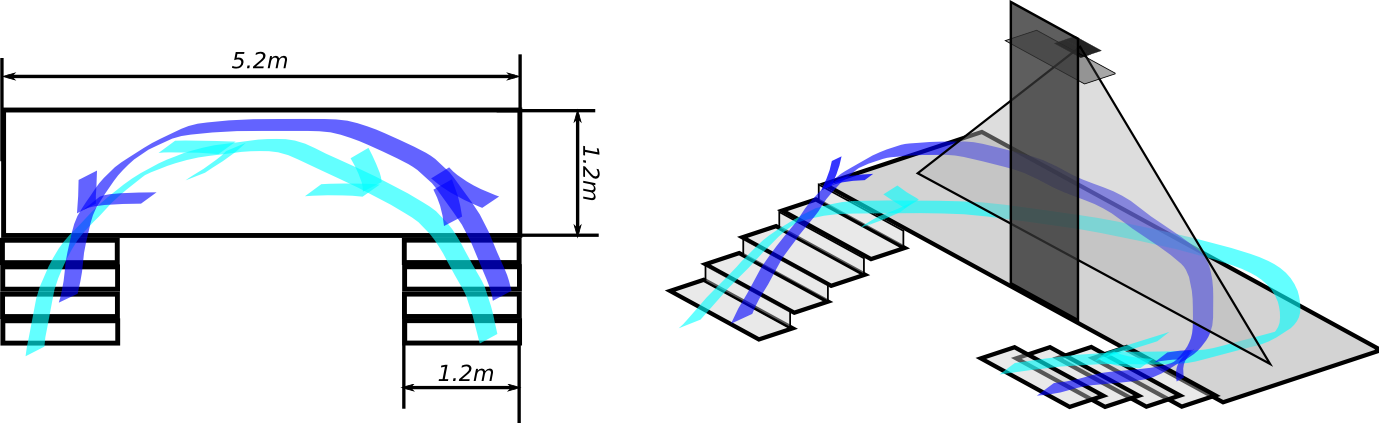}{Schematics of the experimental site; on the
  left: top view; on the right: perspective view. The geometry of the
  staircases at the ends of the landing induces curved
  trajectories.}{expsetup}

It is important to highlight  that our data do not refer to pedestrians instructed \textit{a-priori} to cross the landing (as common in many ``laboratory'' crowd experiments); rather, they refer to the actual, unbiased, ``field'' measurement of pedestrian traffic.

Following an approach similar to the one already introduced by Seer
{\em et al.}~\cite{seer2012kinects}, we performed recordings by employing a standard
commercial \textit{Kinect}\texttrademark\ sensor by Microsoft
Corp.~\cite{Kinect} that allows reliable acquisition of
pedestrian positions.

\begin{figure}[t!]  
\centering 
\includegraphics[width=1\textwidth,trim=9cm 8cm 9cm 8cm,clip]{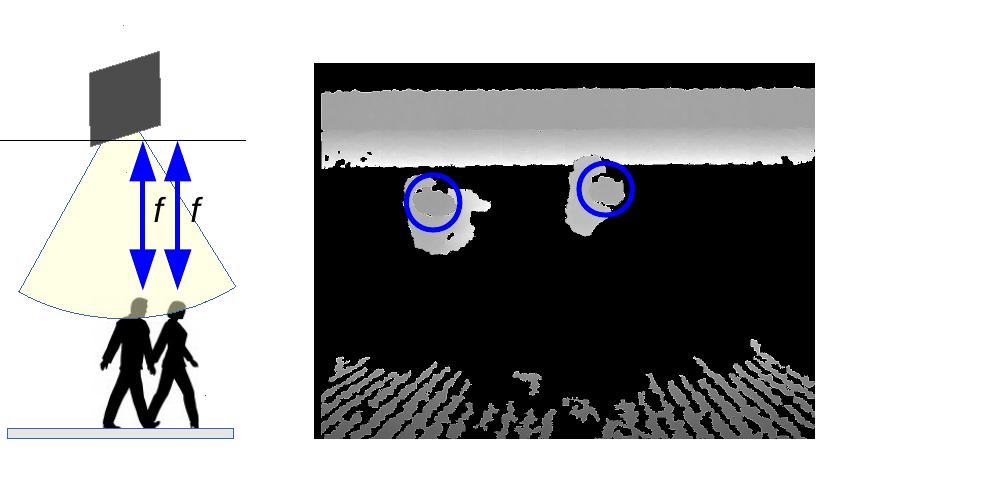} 
\caption{Left: schematic of how the sensor observe pedestrians passing
  in the landing; on the right: actual depth map recorded in the
  measurement site.}
\label{fig-kinect-depth} 
\end{figure}

The Kinect\texttrademark\ sensor is equipped with a special camera
designed to enhance \textit{natural interaction}, i.e.~an interaction
 with computers and the gaming consoles
(Microsoft Xbox 360\texttrademark) based on movements. In particular, in addition to an
ordinary camera, the Kinect\texttrademark\ is able to perform hardware measurements of the \textit{depth map} of the observed scene. The depth map is a two dimensional grey-scale map which associates to every recorded pixel  an intensity proportional to its distance from the camera plane (see
Figure~\ref{fig-kinect-depth}). In our experiments, in order  to track pedestrians, we did not acquire any recordings from the standard camera, rather we  relied totally on the  depth map measurements.


As typically done in literature (cf.~e.g.~\cite{helbing2013pedestrian, DBLP:journals/ijon/BoltesS13, liu2009extraction}) when
one is concerned with the measurement of pedestrian trajectories, it
is often favorable to record the top view of the scene. From the top
view phenomena like partial body exposure to the camera or mutual
hiding, are absent.  The problem of constructing pedestrian
trajectories can be solved via identification and tracking of
heads. Although the heads are always exposed to a top-viewing camera,
performing their automatic tracking without additional hypothesis,
e.g.~on the clothing of pedestrian, can be hard. As pointed out
in~\cite{seer2012kinects}, the depth map associated to the top view of a given
scene can be fruitfully used to detect heads. Heuristically speaking,
the objects which are closest to the camera generate the local minima
in the depth map: such objects, \textit{modulo} a background subtraction, are
\textit{most likely} to be heads.

To generate pedestrian trajectories,  ``raw'' depth maps were first
processed to extract heads positions in each frame; then,
well-established particle tracking algorithms (developed for Particle
Tracking Velocimetry) were used to generate tracks.

\subsection{The algorithm}
We report hereby the overall algorithmic procedure.  Note that some of
the first steps coincide with those explained in~\cite{seer2012kinects}.

Let $f^n=f^n(z)$ be the depth map recorded by Kinect\texttrademark\ (VGA resolution
$640\times 480 px^2$) at time instant $n$ at position $z$, i.e.
\begin{equation}
f^n(z):=dist(\mbox{element in $z$} ,\mbox{camera plane}),
\end{equation}
where $z=(x,y)$ is a point in the VGA frame.
\begin{enumerate}
\item \textbf{Background subtraction.} In the recorded picture, and
  hence in the depth maps, a common background is expected.

To detect pedestrians, the foreground must be first isolated. Let
$B=B(z)$ be a depth map of the background (possibly built after
suitable averages of empty backgrounds), the foreground $F^n=F^n(z)$
associated to a depth map $f^n=f^n(z)$ is obtained through the
thresholding
\begin{equation*}
F^n(z) \leftarrow f^n(z) \cdot [ f^n(z) - B(z) > \epsilon_1 ],
\end{equation*}
where $\epsilon_1>0$ is a given (small) threshold, and $[P] = 1$ if
proposition $``P"$   holds true, $[P] = 0$, otherwise.
\item \textbf{Height thresholding.} A second thresholding operation is
  performed to eliminate objects which, although part of the
  foreground, are too small to be pedestrians, i.e.
\begin{equation*}
F^n(z) \leftarrow F^n(z) \cdot [ F^n(z) > h_1 ].
\end{equation*}
\item \textbf{Generation of a sparse depth map.} For computational
  reasons, the foreground of the thresholded depth map $F^n$ is
  random sampled generating a \textit{sparse} depth map of $N$ samples
\begin{equation*}
F^n_s = \{(z_1,F^n(z_1)),(z_2,F^n(z_2)),\ldots,(z_N,F^n(z_N))\},
\end{equation*}
where every pair $(z_i,F^n(z_i))$ satisfies $F^n(z_i)\neq 0$, i.e.~the
selected sample owns to the foreground and, \textit{likely}, to a pedestrian.
\item \textbf{Sparse samples clusterization and pedestrian detection.}
  In order to identify and isolate pedestrians, sparse samples are
  agglomerated in \textit{macro-samples} which likely are in
  correspondence with pedestrians themselves.

The agglomeration is performed via a \textit{hierarchical clustering}
based on the geometrical distance between points, in particular, a
\textit{complete linkage clustering} algorithm is
used~\cite{duda2012pattern}. Heuristically speaking, the sparse samples get
agglomerated in a binary fashion forming larger and larger
macro-samples. Ideally, macro-samples whose mutual distance is larger
than the scale length of the human body $S$ (e.g., average distance
between the shoulders) do correspond to individuals.

The iterative agglomeration procedure merges macro-samples on the
basis of their distance starting from the closest pairs. In
particular, given two macro-samples $q_1$ and $q_2$, the metric used
satisfies
\[
\left\{
\begin{array}{l}
d^\infty(q_1,q_2) = \max_{(z_1 \in q_1, z_2 \in
  q_2)}d^\infty(z_1,z_2),\\ d^\infty(z_1,z_2) =
d(z_1,z_2),\ \mbox{\textit{if $z_1$ and $z_2$ are \textit{simple}
    samples}},
\end{array}
\right.
\]
where $d$ is the ordinary euclidean distance on the plane.

The agglomeration procedure can be described as reported below. It is
worth to remark that has computational complexity $O(N^3)$, while
optimal computational complexity $O(N^2\log N)$ can be
reached~\cite{duda2012pattern}.

\begin{algorithm}[H]
\label{alg-aggl}
\KwData{Set of all the samples (as macro-samples): $Q_0^n =\{ \{z\}
  \colon x \in F_s^n \}$} $m\leftarrow 0$

\While{ $|Q^n_m| > 1$}{

$(q_1^m,q_2^m) \leftarrow \arg\min_{q'_1,q'_2 \in Q^n_m}
  d^\infty(q'_1,q'_2)$

$Q_{m+1}^n \leftarrow \left(Q^n_{m} - \{q_1^m,q_2^m\}\right) \cup
  \{\{q_1^m,q_2^m\}\} $

$m \leftarrow m+1$
}

\end{algorithm}
The procedure, which iteratively builds the super-sample $Q^n_{N-1}$,
can be visualized via a dendrogram (see Figure~\ref{fig-dendrogram}).

\begin{figure}[t!]  
\centering \includegraphics[width=1.3\textwidth,trim=2cm 14cm 2cm
  2cm,clip]{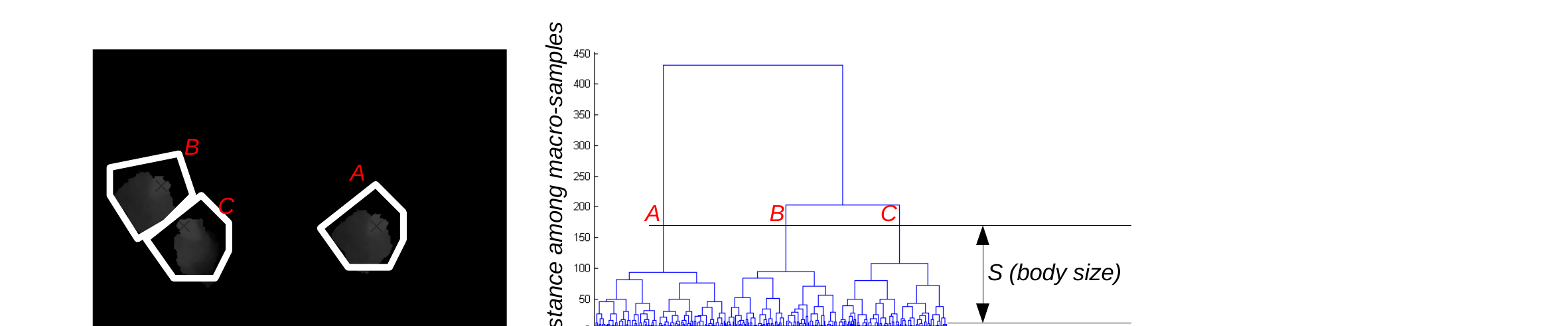}
\caption{Left: depth map in which three pedestrians have been
  separated; on the right: cut dendrogram at height $S$, three
  macro-samples are present.}
\label{fig-dendrogram} 
\end{figure} 

To conclude, we observe that if the agglomeration procedure is
truncated at a step $m$ such that
\begin{equation*}
\left\{
\begin{array}{l}
d^\infty(q_1^{m},q^{m}_2) < S\\ d^\infty(q_1^{m+1},q^{m+1}_2) \geq S,
\end{array}
\right.
\end{equation*}
then the super-samples in $Q_m^n$ feature a mutual distance larger than
$S$; as such, they are associated with pedestrians.

\item \textbf{Head position estimation.} Given pedestrian identified
  by the cluster $C_j^n \in Q_m^n$, for some $j$, $1 \leq j\leq N^n$ (being $N^n$ the total number of pedestrians present in frame $n$),  we consider as head the set $H_j^n\subset C^n_j$ of samples whose depth is smaller than a given percentile $\alpha_k$  (usually  $k=10$) of the depth distribution of $C_j^n$, 
\begin{equation*}
H_j^n = \{ z \in C_j^n \colon \mbox{\textit{depth}}(z) \geq \alpha_k \}.
\end{equation*}
The head position is then estimated considering the centroid of the set, in formulas
\begin{equation*}
\bar{z}^{n}_j = (x_j^n,y_j^n) = \mbox{\textit{mean}}(H_j^n).
\end{equation*}    

\item \textbf{Linking positions across the frames.} An estimate $\bar{z}^{n}_j$ of 
  the heads position across the frames has been given. In order to
  approximate the pedestrian trajectories, heads positions must hence
  be tracked.  

The problem of tracking time-sampled particle positions has been
studied in several fields, in particular, it is central in
Experimental Fluid Dynamics when a Lagrangian approach to flows is
pursued. Following standard approaches in experimental Particle
Tracking Velocimetry (PTV), and especially via
OpenPTV~\cite{willneff2003spatio,openPTV}, Pedestrian trajectories are generated. In particular, sequences in the form of Equation~\eqref{eq-trajectory-single} are obtained.
\item \textbf{Estimation of kinematic observables associated to
  trajectories.} Once the trajectories are known in their sampled form,
  kinematic observables such as velocities and accelerations must be
  estimated. Since the head estimation procedure is not exempt of
  measurement noise, and a certain degree of regularity in
  trajectories is expected, a smoothing spline (with smoothing
  parameter $\lambda=1$) is used~\cite{de1978practical} (See
  Figure~\ref{fig-tracks} for an example of some of the final
  trajectories).

\begin{figure}[t!]  
\centering \includegraphics[width=1.5\textwidth,trim=0cm 10.5cm 4cm
  6cm,clip]{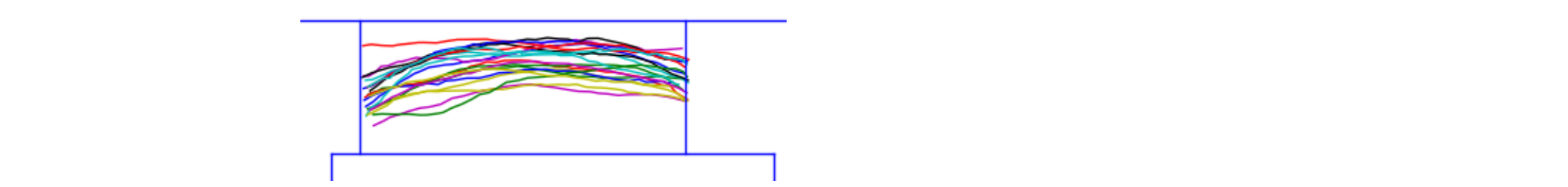}
\caption{A sample selection of 20 pedestrian trajectories in the final
  data.}
\label{fig-tracks} 
\end{figure} 
\end{enumerate}
After such procedure, the set of trajectories including relative velocities and accelerations is deducted.

\section*{Acknowledgments} 
We would like to thank Ad Holten and Gerald Oerlemans for the help
with the installation of the Kinect\texttrademark\ sensor in the MetaForum building
at TU/e. We thank Dr.~Alex Liberzon (Tel Aviv, Israel) for his
precious help with the adaption of the Particle Tracking software to
our project.  We acknowledge the hosting of Lorentz Center (Leiden,
The Netherlands), where, during the workshop ``Modeling with Measures'', a part of this paper has been written.
KV would like to acknowledge the support of NWO VICI grant 639.033.008.

\bibliographystyle{AIMS} 
\bibliography{master}

\medskip
\medskip

\end{document}